\documentclass[preprint,floatfix,superscriptaddress,nofootinbib]{revtex4-2}
\usepackage{booktabs} 
\usepackage{dcolumn}

\usepackage[export]{adjustbox} 

\usepackage[table]{xcolor}
\usepackage{makecell}

\draft 

\usepackage{threeparttable}
\usepackage{graphicx}

\begin{document}

\title{Quantum Calculations of Hydrogen Absorption and Diffusivity in Bulk $\mathrm{CeO_2}$} 
\author{Jared C. Stimac }
\email[]{stimac1@llnl.gov}
\affiliation{Lawrence Livermore National Laboratory, Livermore, CA, 94550, USA.}

\author{Nir Goldman }
\affiliation{Lawrence Livermore National Laboratory, Livermore, CA, 94550, USA.}
\affiliation{Department of Chemical Engineering, University of California, Davis, CA, 95616, USA.}

\begin{abstract}
CeO$_2$ (ceria) is an attractive material for heterogeneous catalysis applications involving hydrogen due to its favorable redox activity combined with its relative impermeability to hydrogen ions and molecules. However, to date, many bulk ceria/hydrogen properties remain unresolved in part due to a scarcity of experimental data combined with quantum calculation results that vary according to the approach used. In this regard, we have conducted a series of Density Functional Theory (DFT) calculations utilizing generalized gradient (GGA), meta-GGA, and hybrid functionals as well as several corrections for electronic correlations, applied to a number of properties regarding hydrogen in bulk stoichiometic $\mathrm{CeO_2}$. Our calculations place reasonable bounds on the lattice constants, band gaps, hydrogen absorption energies, and O-H bond vibrational frequencies that can be determined by DFT. In addition, our results indicate that the activation energy barriers for hydrogen bulk diffusion are uniformly low ($ < 0.15 \ \mathrm{eV} $) for the calculation parameters probed here and that, in general, the effect of hydrogen tunneling is small at ambient temperatures. Our study provides a recipe to determine fundamental physical chemical properties of Ce-O-H interactions while also determining realistic ranges for diffusion kinetics. This can facilitate the determination of future coarse-grained models that will be able to guide and elucidate experimental efforts in this area. 

\end{abstract}


\maketitle 

\section{Introduction}
In the last several decades, ceria ($\mathrm{CeO_2}$) has been the subject of numerous experimental and theoretical studies because of its potential in an array of applications predominantly in heterogeneous catalysis, including water-gas shift reactions\cite{shido1992regulation, panagiotopoulou2007water}, water splitting\cite{sun2020boosting}, and semi-hydrogenation of alkynes\cite{garcia2014unique}. However, the mechanism for hydrogen absorption and transport in ceria remains an open question. Chemisorption of hydrogen ions resulting from H$_2$ dissociation or H$_2$O splitting is exothermic and forms hydroxyl species on the surface, which could act as a first step in the formation of cerium hydrides.\cite{li2021interaction, werner2017toward,schweke2022cerium,schweke2022elucidating} In contrast, hydrogen can require relatively high temperatures and pressures to diffuse into the bulk of sub-stoichiometric surfaces, where it can form hydroxyl groups and hydride ions near oxygen vacancies.\cite{li2021interaction, mao2024hydrogen} Such studies would benefit from atomistic calculations that could help determine initial hydriding reaction steps as well as the ensuing chemical rate parameters, which would help elucidate the importance of competing chemical mechanisms.

In this regard, quantum calculations with Kohn-Sham Density Functional Theory (DFT) remains a popular choice for determining the breaking and forming of bonds in f-electron systems (e.g., Refs.~\citenum{Goldman17,Goldman_U_H_2022,Goldman_PuO2H_2022}). DFT calculations regarding ceria surface chemistry, though, largely remain quantitatively inconsistent.
There has been reported a wide range of energetic barriers for diffusion into the bulk from $\mathrm{CeO_2}$(111) surfaces, with values between $1.15 \ \mathrm{eV}$ and $1.67 \ \mathrm{eV}$ (Refs.~\citenum{wu2015role} and \citenum{marrocchelli2012first}, respectively). In addition, there exist similar discrepancies in DFT data regarding hydrogen bulk diffusion in stoichimetric CeO$_2$, with results ranging from $0.18 \ \mathrm{eV}$ in Ref.~\citenum{wu2015role} to $0.52 \ \mathrm{eV}$ in Ref.~\citenum{marrocchelli2012first}.
This is in sharp contrast to one set of experimental results\cite{mao2024hydrogen} utilizing nuclear reaction analysis (NRA) on sub-stoichiometric $\mathrm{CeO_{1.69}}$ films exposed to $\mathrm{H_2}$  to estimate a diffusion activation barrier of $1.69 \ \mathrm{eV}$.

The range of DFT results as well as the seeming disagreement with experimental diffusion barriers is likely in part due to the sensitivity of H-Ce-O interactions to different calculation parameters, including choice of exchange-correlation functional and level of theory (discussed below) and use of specific Hubbard parameters (e.g., DFT+U) to better account for electron correlations. For example, while the Hubbard U correction is generally only applied to the Ce 4$f$ orbitals, some data on a number of metal oxides (including ceria) indicate advantages to including additional Hubbard corrections to the O 2$p$ orbitals to improve the description of defect states.\cite{nolan2005electronic, nolan2006hole, may2020improved,keating2012analysis, plata2012communication} Thus, to the best of our knowledge, there does not exist a systematic determination of how hydrogen properties within CeO$_2$ depend on the various options available here, and how different choices might affect the interpretation of results and possible comparison to experiments. 

In this work, we address these issues by employing a range of DFT calculations at various levels of theory to better elucidate $\mathrm{CeO_2}$/H interactions. We report on calculations with the generalized gradient (GGA), meta-GGA, and screened hybrid functionals, using a Hubbard U correction on Ce 4$f$ orbitals as well as a wide range of values for the additional correction to O 2$p$. We then report on bulk stoichiometric $\mathrm{CeO_2}$ properties, interstitial hydrogen formation energies, and diffusion activation energy barriers. 
We include Arrhenius prefactor values and O--H bond vibrational frequencies, where applicable. Finally, we use the characteristic crossover temperatures to estimate the significance of quantum  nuclear tunneling effects vs. classical, over-the-barrier hopping mechanisms.
We believe our results place accurate bounds on hydrogen diffusion within bulk $\mathrm{CeO_2}$ that can help elucidate experimental results while also providing guidance for future DFT studies on similar systems.

\section{Methods}
\label{sec:Methods}
All DFT calculations were performed using the Vienna \textit{ab initio} Simulation Package (VASP) \cite{kresse1993vasp1, kresse1996vasp2, kresse1996vasp3} with Projector Augmented Wave (PAW) pseudopotentials \cite{blochl1994projector, kresse1999ultrasoft}. For our studies we have used the following exchange correlation functionals: the Perdew-Burke-Ernzerhov (PBE)\cite{perdew1996generalized} and the revised PBE for solids (PBEsol)\cite{perdew2008restoring} generalized gradient (GGA) functionals, the strongly constrained and appropriately normed (SCAN)\cite{sun2015strongly} meta-GGA functional, and the Heyd-Scyseria-Ernzerof (HSE06) \cite{krukau2006influence} and HSEsol\cite{schimka2011improved} hybrid functionals. The D3 dispersion correction \cite{grimme2010consistent} was used in all calculations except for those with HSEsol, for which we were unable to find a D3 parameterization. 

In order to account for the self-interaction in non-hybrid functionals, we use the rotationally invariant DFT+U formalism by Dudarev et al.\cite{dudarev1998electron}, where $\mathrm{U_{eff} = U - J}$. We denote particular combinations of $\mathrm{U_{eff}}$ values discussed by listing the Ce 4\textit{f} ($\mathrm{U^{Ce_{4f}}_{eff}}$) and O 2\textit{p} ($\mathrm{U^{O_{2p}}_{eff}}$) values in curly brackets in units of eV next to the functional name. For example, a calculation using PBE with $\mathrm{U_{eff}}$ values of  $ 6 \ \mathrm{eV}$ on the Ce 4\textit{f} interactions and $0 \ \mathrm{eV}$ on the O 2\textit{p} interactions is written as PBE$\{6,\ 0\}$. In general, for non-hybrid functional calculations the $\mathrm{U_{eff}}$ values on the Ce $f$ orbitals is set to $6 \ \mathrm{eV}$. Curly brackets were omitted for results from the hybrid functionals HSE06 and HSEsol since these already take the electronic self-interaction into account and no Hubbard $\mathrm{U}$ value is necessary.  

Previous studies have reported a non-magnetic ground state of perfect crystal $\mathrm{CeO_2}$, which we have confirmed with for all functional/$\mathrm{U_{eff}}$ combinations in this work. However, due the potential for defect-induced magnetization,\cite{han2009oxygen, keating2009intrinsic, keating2012analysis} we have conducted spin-polarization on interstitial hydrogen in bulk $\mathrm{CeO_2}$ which retain a nonmagnetic state. Hence, all calculations reported here were performed without spin-polarization.

Lattice optimizations were carried out via the conjugate gradient algorithm with a maximum-force convergence criterion of $0.01 \ \mathrm{eV/\AA}$ using the four formula unit (CeO$_2$) cubic supercell, a Mohnkhorst-Pack\cite{monkhorst1976special} $6\times6\times6$ k-point mesh, and a $500 \ \mathrm{eV}$ plane-wave cutoff. Calculations including interstitial hydrogen were performed with a 32 formula unit cubic super cell (97 atoms total), Mohnkhorst-Pack\cite{monkhorst1976special} k-point mesh of $2\times2\times2$, and a $500 \ \mathrm{eV}$ plane-wave cutoff. 
Gaussian smearing of the electrons with a value of $0.20 \ \mathrm{eV}$ was used throughout. 
Additional results related to numerical convergence are given in Appendix 
\ref{sec:Convergence}.

\section{Results and Discussion}
\subsection{Bulk $\mathrm{CeO_2}$ properties}
We now assess the computed bulk properties of perfect crystals $\mathrm{CeO_2}$ for the array of functionals and values of $\mathrm{U^{O_{2p}}_{eff}}$ in our study, with comparison to the lattice parameter, bulk modulus, and electronic band gaps (Table~\ref{table:property_table}). Here, we make comparison to the zero temperature experimental lattice parameter ($a_0$) determined by extrapolation from room temperature using the thermal expansion coefficient.\cite{castleton2007} While thermal effects on $a_0$ are small ($< 1\%$), these remain pertinent given the relatively small differences between DFT results. Experimental results at both room temperature and 0~K are shown in Table~\ref{table:property_table} for the sake of completeness.

We observe that the most accurate non-hybrid functional value of $a_0$ comes from PBEsol$\{6,\ 12\}$ (5.446 \AA), which differs from the experimental value (5.391 \AA) by $\sim$1.0\%.  HSE06 (5.382 \AA) yields an error of $\sim0.2 \%$, and HSEsol (5.357 \AA) yields an error of $\sim0.6\%$.  The largest deviations were observed from PBE$\{6,\ 0\}$ (5.482 \AA) with an error of $\sim 1.6 \%$. The use of $\mathrm{U^{O_{2p}}_{eff}}$ yields a small but systematic shift in $a_0$, where PBE, PBEsol, and SCAN all yields values that improve to agreement with experiment of $\sim 0.04$~\AA\ upon increasing its value from zero to twelve.

Next, we consider results for the two commonly measured band gaps of CeO$_2$. The O2$p$ to Ce4$f$ band gap is underestimated by all of our DFT results excluding those from HSE06 and SCAN$\{6,\ 12\}$, both of which yield a value of 3.0~eV, which agrees with the experimental value from Ref.~\citenum{castleton2007}. However, we note that results from HSEsol, PBEsol$\{6,\ 12\}$, and all of our SCAN calculations are within the experimental confidence interval of $\pm$0.25~eV from Ref.~\citenum{castleton2007}. In contrast, we find that the results for the O2$p$ to Ce5$d$ band gap from HSE06 is too large, with a value of 6.8~eV, compared to the experimental values of 5.75~eV (Ref.~\citenum{castleton2007}) and 6.0~eV (Ref.~\citenum{wuilloud1984spectroscopic}).\cite{sun2019first} Non-zero values of $\mathrm{U^{O_{2p}}_{eff}}$ improve the agreement with experiment, overall.
Results from PBE$\{6,\ 12\}$ and SCAN$\{6,\ 12\}$ give the best agreement , with values of 5.8 and 5.9~eV, respectively. In addition, the results from PBE$\{6,\ 6\}$, PBE$\{6,\ 12\}$, PBEsol$\{6,\ 12\}$, SCAN$\{6,\ 6\}$, and SCAN$\{6,\ 12\}$ are all within the experimental confidence interval. 
Otherwise, the remaining results from PBE and SCAN, and all of the results from PBEsol all underestimate the O2$p$ to Ce5$d$ band gap by $\sim$ 0.45~eV.

Finally, we compare results for the CeO$_2$ bulk modulus ($K$) to the range of available experimental results.\cite{nakajima1994defect, gerward2005bulk, gerward1993powder, duclos1988high} The bulk modulus was computed by fitting energy-volume data at eight uniformly sampled values about the equilibrium lattice parameter $a_0 \ \pm 7\% \ $, using the optimized four formula unit cell. The data was then fit to the Birch-Murnaghan\cite{birch1947finite} equation of state as implemented in the Atomic Simulation Environment (ASE)\cite{larsen2017atomic} python package. HSE functionals were not used in this part of our study due to their computational expense. We observe that all values from PBEsol and SCAN in our study fall within the range of experimental values of 204 - 236~GPa. On the other hand, results from PBE were consistently too low, with PBE$\{6,\ 12\}$ yielding the closest agreement with a value of 195.9~GPa. In all cases, inclusion of a non-zero value of $\mathrm{U^{O_{2p}}_{eff}}$ increases the value of $K$, as expected, though the degree of increase is relatively small. For example, increasing $\mathrm{U^{O_{2p}}_{eff}}$ from zero to 12 for PBEsol and SCAN results in an increase in $\mathrm{K}$ by $< 10 \ \mathrm{GPa}$, or $\sim 5\%$.

Overall, we find that for bulk, defect-free CeO$_2$, use of $\mathrm{U^{O_{2p}}_{eff}} = 12$ generally improved the accuracy of our calculations. In particular, PBEsol$\{6,\ 12\}$ and SCAN$\{6,\ 12\}$ were the most consistently accurate calculation sets employed in our study, even in comparison to HSE06 and HSEsol. However, results from PBE overestimate the lattice constant and underestimated the electronic band gaps and bulk modulus. Hence, non-zero values of $\mathrm{U^{O_{2p}}_{eff}}$ appear to have some utility, based on this initial part of our study.

\begin{table}
    \caption{Lattice parameter $a_0$, band gaps, and the bulk modulus $K$ from various functionals and values of $\mathrm{U_{eff}}$. The color shading surrounding entries corresponds to the absolute error normalized by the maximum error for each property such that darker shaded values have smaller absolute error. Errors for the lattice parameter and band gaps were evaluated using values reported by Castelton et. al.\cite{castleton2007}. All values of $K$ within the experimental range of 204 - 236~GPa are shaded to the same degree.}
    \begin{center}
    

\begin{ruledtabular}
\begin{tabular}{cccccccc}
      & \multicolumn{2}{c}{$\mathrm{U_{eff}}$ [eV]}         &                & \multicolumn{2}{c}{Band gap [eV]}                 &       \\
$\mathrm{E_{XC}}$& Ce 4\textit{f}    &  O 2\textit{p}       &  $a_0$ \ [\AA] & O2\textit{p}$\rightarrow$Ce4\textit{f} & O2\textit{p}$\rightarrow$Ce5\textit{d} & $K$ [GPa] \\
\hline
PBE     & 6 &            0 &  \cellcolor{blue!0.0}    5.482 &  \cellcolor{blue!0} 2.5  &  \cellcolor{blue!18.8} 5.3 &                       187.4 \\
        & 6 &            6 &  \cellcolor{blue!8.4}    5.463 &  \cellcolor{blue!8} 2.6  &  \cellcolor{blue!28.2} 5.5 &                       191.3 \\
        & 6 &           12 &  \cellcolor{blue!15.8}   5.446 &  \cellcolor{blue!0} 2.5  &  \cellcolor{blue!37.6} 5.8 &                       195.9 \\
PBEsol  & 6 &            0 &  \cellcolor{blue!23.3}   5.429 &  \cellcolor{blue!8} 2.6  &  \cellcolor{blue!18.8} 5.3 &   \cellcolor{blue!40} 203.9 \\
        & 6 &            6 &  \cellcolor{blue!31.2}   5.411 &  \cellcolor{blue!8} 2.6  &  \cellcolor{blue!18.8} 5.3 &   \cellcolor{blue!40} 207.8 \\
        & 6 &           12 &  \cellcolor{blue!39.1}   5.389 &  \cellcolor{blue!24} 2.8 &  \cellcolor{blue!28.2} 5.5 &   \cellcolor{blue!40} 212.4 \\
SCAN    & 6 &            0 &  \cellcolor{blue!8.4}    5.463 &  \cellcolor{blue!24} 2.8 &  \cellcolor{blue!21.6} 5.4 &   \cellcolor{blue!40} 206.8 \\
        & 6 &            6 &  \cellcolor{blue!16.7}   5.444 &  \cellcolor{blue!24} 2.8 &  \cellcolor{blue!32.9} 5.6 &   \cellcolor{blue!40} 211.1 \\
        & 6 &           12 &  \cellcolor{blue!26.4}   5.422 &  \cellcolor{blue!40} 3.0 &  \cellcolor{blue!32.9} 5.9 &   \cellcolor{blue!40} 216.3 \\
HSE06   & 0 &            0 &  \cellcolor{blue!36.0}   5.382 &  \cellcolor{blue!40} 3.0 &  \cellcolor{blue!0.00} 6.8 &             - \\
HSEsol   & 0 &           0 &  \cellcolor{blue!26.1}   5.357 &  \cellcolor{blue!24.8} 2.8 &  \cellcolor{blue!19.7} 6.2 &            - \\

Experiment &           - &            - &      \makecell{5.411\cite{gy1998binary} \\ 5.391(Extrap. to 0 K)\cite{castleton2007}} &             \makecell{3.3\cite{wuilloud1984spectroscopic} \\3.0$\pm 0.25$ \cite{castleton2007}} &           \makecell{6.0\cite{wuilloud1984spectroscopic} \\  5.75 $\pm 0.25$ \cite{castleton2007}}&      204 - 236 \cite{nakajima1994defect, gerward2005bulk, gerward1993powder, duclos1988high} \\
\end{tabular} 
\end{ruledtabular} \label{table:property_table}
    
    \end{center}
\end{table}

\subsection{Interstitial hydrogen formation energies}
\begin{figure}
    \includegraphics{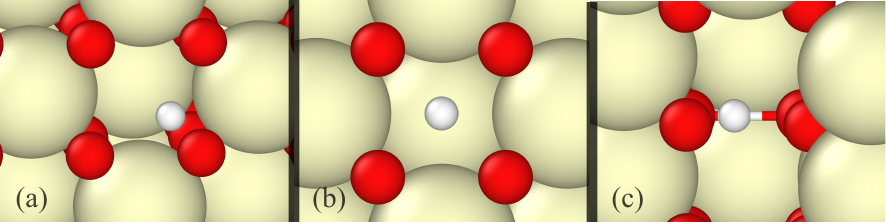}
    \caption{\label{fig:Energy_min_configs} Three energy minima configurations: (a) int, (b) oct, and (c) lin. }
\end{figure}

\begin{table}
    \caption{Hydrogen interstitial formation energies. The color shading surrounding entries corresponds to the absolute error (HSE06 as the ground truth) normalized by the maximum error for each energy such that darker shaded values have smaller absolute error. }
    \begin{center}
    \begin{ruledtabular}
\begin{tabular}{cccccc}

      & \multicolumn{2}{c}{$\mathrm{U_{eff}}$ [eV]}         &              &             &           \\[-.1cm]
$\mathrm{E_{XC}}$& Ce 4\textit{f}    &  O 2\textit{p}       &  int [eV] & oct [eV] & lin [eV] \\
\hline
PBE     & 6 &        0 &  \cellcolor{blue!11.7}   1.09   & \cellcolor{blue!4.59}  2.49   &  \cellcolor{blue!9.60}  1.04   \\
        & 6 &        6 &  \cellcolor{blue!6.33}  1.02    &  \cellcolor{blue!7.58} 2.56  &  \cellcolor{blue!2.09}  0.96   \\
        & 6 &       12 &  \cellcolor{blue!0.00}  0.94    &  \cellcolor{blue!11.4} 2.65   &  \cellcolor{blue!0.00} 0.94   \\
PBEsol  & 6 &        0 &  \cellcolor{blue!21.2}  1.20    & \cellcolor{blue!0.00}  2.38   &  \cellcolor{blue!18.5}  1.14   \\
        & 6 &        6 &  \cellcolor{blue!14.8}  1.12    &  \cellcolor{blue!2.38} 2.43   &  \cellcolor{blue!13.1} 1.08   \\
        & 6 &       12 &  \cellcolor{blue!9.77}  1.06    &  \cellcolor{blue!5.54} 2.51   &  \cellcolor{blue!6.24} 1.01   \\
SCAN    & 6 &        0 &  \cellcolor{blue!35.1}  1.37    &  \cellcolor{blue!16.5} 2.76   &  \cellcolor{blue!38.5} 1.37   \\
        & 6 &        6 &  \cellcolor{blue!30.7}  1.32    &  \cellcolor{blue!20.2} 2.85   &  \cellcolor{blue!35.3} 1.33   \\
        & 6 &       12 &  \cellcolor{blue!25.7}  1.26    &  \cellcolor{blue!25.3} 2.97   &  \cellcolor{blue!31.2}  1.29   \\
HSE06\footnotemark[1] & 0 &        0 &    1.43    &   3.31       &   1.39       \\
HSEsol\footnotemark[1]\footnotemark[2]  & 0 &        0 &    1.36    &   3.19       &   1.31       \\

\end{tabular}
\end{ruledtabular}
\footnotetext[1]{Estimated with fixed geometry obtained from SCAN\{6, 12\} relaxation}
\footnotetext[2]{No D3 correction}
    \end{center}
    \label{table:Formation_energies}
\end{table}

Next, we consider the hydrogen interstitial formation energies in bulk $\mathrm{CeO_2}$. We have identified three energetic minima, which are displayed in Figure \ref{fig:Energy_min_configs}. The ``interstitial'' site (labeled \emph{int}) consists of a hydroxyl (OH) species, with the hydrogen atom bonded to a single oxygen with a bond length of $\sim 1.0 \ \mathrm{\AA}$.
The octahedral site (labeled \emph{oct}) corresponds to a hydrogen ion at the center of one of the octahedra resulting from the Ce face-centered cubic sub-lattice, involving six equidistant neighboring Ce atoms. 
We note that prior studies\cite{chafi2009density, sohlberg2001interactions} indicate that the oct site was unstable and would relax to the int site upon energy minimization of the atomic positions, though these efforts used GGA functionals without any Hubbard U correction. 

An additional absorption minima was found by rotating the int site O--H bond to be in line with another neighboring oxygen so that the hydrogen bisects the O--O axis. We refer to this new site as the \emph{lin} (e.g., linear) site. Upon rotation towards a neighboring oxygen site, the O--H bond stretches due to the formation of a hydrogen-bond to the neighboring oxygen (i.e., O--H...O). Hydrogen-bonding within the substrate has been observed in other $\mathrm{CeO_2/H}$ DFT studies,\cite{chafi2009density} as well as other metal oxides.\cite{gordon2014hydrogen, sundell2007density, bjorketun2007effect, hermet2013hydrogen}

The formation energies for all interstitial configurations were determined using the equation:  
\begin{equation}
E_\mathrm{Form} = E(\mathrm{CeO_2/H}) - N_\mathrm{CeO_2} E(\mathrm{CeO_2}) - \frac{1}{2}E(\mathrm{H_2})
\end{equation}

\noindent where $E(\mathrm{CeO_2/H})$ is the total energy of the $\mathrm{CeO_2/H}$ system, $E(\mathrm{CeO_2})$ is the energy per CeO$_2$ formula unit, $N_\mathrm{CeO_2}$ is the number of formula units in the system in question, and $E(\mathrm{H_2})$ is the molecular hydrogen energy. Once again, the formation energies from HSE06 and HSEsol were estimated by single point energy calculations using the optimized geometry from SCAN$\{6,\ 12\}$ scaled to their respective lattice dimensions. 

We observe that PBE, PBEsol, and SCAN all underpredict the int site formation energy relative to the HSE06 and HSESol estimated values of 1.43 and 1.36~eV, respectively, by up to several tenths of an eV (Table~\ref{table:Formation_energies}). In addition, use of $\mathrm{U^{O_{2p}}_{eff}}$ caused the formation energy to become smaller. PBE yielded the lowest results with a value of 1.09~eV for PBE$\{6,\ 0\}$ which monotonically decreases to 0.94~eV for PBE $\{6,\ 12\}$. 
We computed a value of 1.20~eV from PBEsol$\{6,\ 0\}$, which decreased to 1.12~eV for PBEsol$\{6,\ 6\}$ and further to 1.06~eV for PBEsol$\{6,\ 12\}$. 
The SCAN set was closest overall to the HSE functionals, with a highest value of 1.37~eV for SCAN$\{6,\ 0\}$ and lowest value of 1.26~eV for SCAN$\{6,\ 12\}$.
We compute the lin site formation energy to be nearly iso-energetic with the int site and with similar a trend, where results from all functionals and $\mathrm{U^{O_{2p}}_{eff}}$ values were up to 0.07~eV lower (less positive) than the int result.
In all cases, non-zero $\mathrm{U^{O_{2p}}_{eff}}$ values caused the formation energy to decrease. 

Results for the oct formation energy show a larger spread relative to results from HSE06 (3.31~eV) and HSESol (3.19~eV), and with an opposite trend for $\mathrm{U^{O_{2p}}_{eff}}$. PBE$\{6,\ 0\}$ yields a value 2.49~eV (0.82~eV lower than HSE06) and PBEsol$\{6,\ 0\}$ a value of 2.38~eV (0.92~eV lower) and SCAN$\{6,\ 0\}$ a value 0f 2.76~eV (0.55~eV lower). By comparison, the energy increases by to 2.65~eV for PBE$\{6,\ 12\}$ (net increase of 0.16~eV), 2.51~eV for PBEsol$\{6,\ 12\}$ (net increase of 0.13~eV), and 2.97~eV for SCAN$\{6,\ 12\}$ (net increase of 0.23~eV), which brings the SCAN result to within 0.34~eV of HSE06. 

\subsection{O--H bond vibrational frequency}

\begin{table}
    \caption{O--H bond frequency as determined from highest frequency calculated at the int site. The color shading surrounding entries corresponds to the absolute error normalized by the maximum error (such that darker shaded values have smaller absolute error)}
    \begin{center}
    \begin{ruledtabular}
\begin{tabular}{cccc}
                 & \multicolumn{2}{c}{$\mathrm{U_{eff}}$ [eV]}         &   \\[-.1cm]
$\mathrm{E_{XC}}$& Ce 4\textit{f}    &  O 2\textit{p}       &   $\nu \ \mathrm{[cm^{-1}]}$   \\
\hline
PBE     & 6 &        0 & \cellcolor{blue!33.8}  3490   \\
        & 6 &        6 & \cellcolor{blue!36.5}  3499   \\
        & 6 &       12 & \cellcolor{blue!33.6}  3490   \\
PBEsol  & 6 &        0 & \cellcolor{blue!9.53}  3413   \\
        & 6 &        6 & \cellcolor{blue!12.9}  3424   \\
        & 6 &       12 & \cellcolor{blue!9.91}  3414   \\
SCAN    & 6 &        0 & \cellcolor{blue!7.70}  3613   \\
        & 6 &        6 & \cellcolor{blue!0.00}  3637   \\
        & 6 &       12 & \cellcolor{blue!2.11}  3631   \\
Experiment & - &     - &    3510\cite{badri1996ftir}   \\

\end{tabular}
\end{ruledtabular}
    \end{center}
    \label{table:OH_frequencies}
\end{table}

We now benchmark our DFT calculations for the predicted O--H vibrational frequency ($\nu \mathrm{(OH)}$) at the int site (Table~\ref{table:OH_frequencies}), which is taken to be the highest frequency from dynamical matrix calculations. Here, the normal mode harmonic frequencies for all interstitial hydrogen minima were obtained with displacement magnitude of $10^{-3} \ \mathrm{\AA}$. Once again, dynamical matrix calculations were not performed with either hybrid functionals due to their extreme computationally expense. 
We observe the largest error, relative to one experimental value\cite{badri1996ftir} of $3510 \ \mathrm{cm^{-1}}$, from SCAN$\{6,\ 6\}$, with a difference of 127~cm$^{-1}$, or a scale factor of $\sim$1.036. PBE $\{6,\ 6\}$ yields the closest result with an error of only 11~cm$^{-1}$, or a scale factor of $\sim$0.997. We note that corrective frequency scaling factors are generally $0.90 - 0.99$, depending on functional and basis set\cite{irikura2005uncertainties}. We thus find that all predicted bond vibrations are within the range of uncertainty typical of \textit{ab initio} computed results. Overall, the O--H vibrational frequency predicted by PBE$\{6,\ 6\}$ is closest to the experiment with a value of 3499 cm$^{-1}$, and the effect of $\mathrm{U^{O_{2p}}_{eff}}$ is mostly insignificant, causing changes on the order of $10 - 20$ cm$^{-1}$. 

\subsection{Kinetic pathways of H bulk diffusion}
\begin{figure}
    \includegraphics{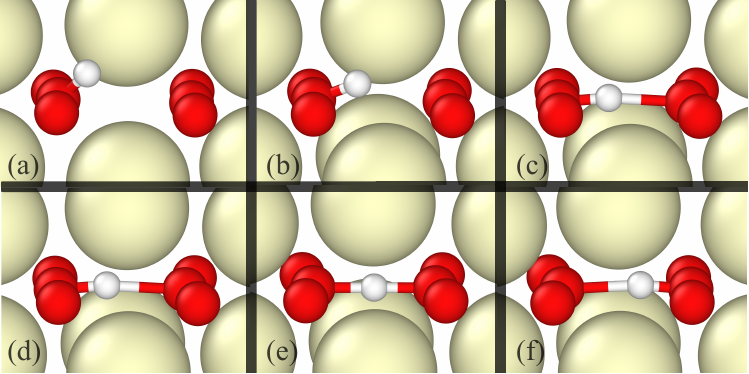}
    \caption{\label{fig:int_int_configs} Critical point configurations for int-to-lin (top row) and lin-to-lin (bottom row) MEPs. Subfigures (a) - (c) and (d) - (f) correspond to configurations 1 through 3 and 3 through 5 in Fig.~\ref{fig:int_int_rxn_coord}, respectively.}
\end{figure}

\begin{figure}
\includegraphics{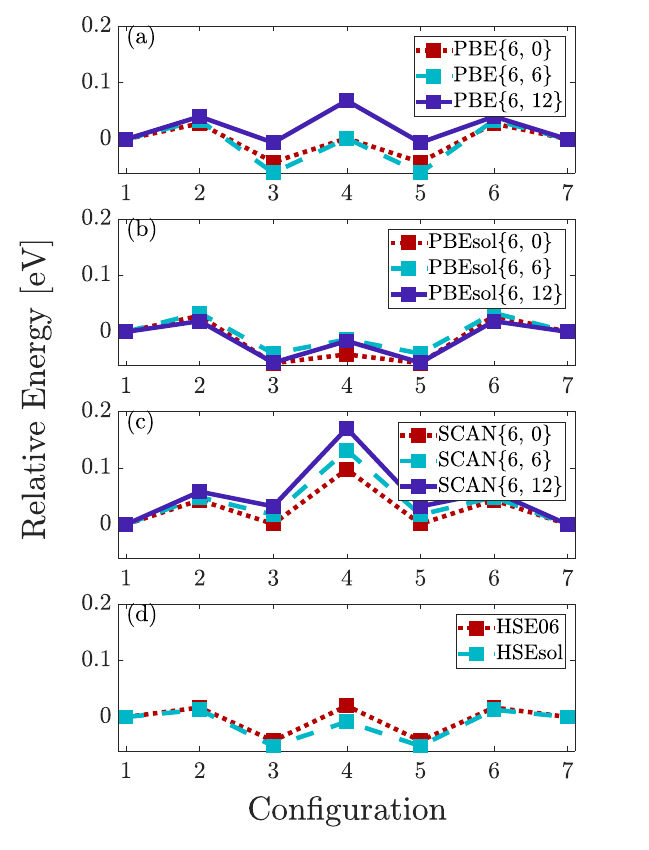}
\caption{\label{fig:int_int_rxn_coord} Reaction coordinate of H diffusing between two neighboring int sites. Energies are relative to the int site energy}
\end{figure}

\begin{table}
    \caption{Transition energy barriers $\Delta E$ and crossover temperatures $T_{\mathrm{c}}$ for barriers between neighboring int sites.}
    \begin{center}
    \begin{ruledtabular}
\begin{tabular}{ccccccc}
    & \multicolumn{2}{c}{$\mathrm{U_{eff}}$ [eV]}  & \multicolumn{2}{c}{int-lin}  &   \multicolumn{2}{c}{lin-lin}              \\[-.1cm]
$\mathrm{E_{XC}}$& Ce 4\textit{f}    &  O 2\textit{p}      &  $\Delta E$ [eV] & $T_\mathrm{c}$ [K] & $\Delta E$ [eV] & $T_\mathrm{c}$ [K] \\
\hline
 PBE     & 6 &        0 &   0.03       &  52 & 0.04  &    138  \\
         & 6 &        6 &   0.03       &  52 & 0.06  &    187  \\
         & 6 &       12 &   0.04       &  56 & 0.07  &    208 \\
 PBEsol  & 6 &        0 &   0.03       &  57 & 0.02  &     80 \\
         & 6 &        6 &   0.03       &  54 & 0.02  &    113 \\
         & 6 &       12 &   0.02       &  56 & 0.04  &    151 \\
 SCAN    & 6 &        0 &   0.04       &  54 & 0.10  &    284 \\
         & 6 &        6 &   0.05       &  50 & 0.11  &    313 \\
         & 6 &       12 &   0.06       &  58 & 0.14  &    348 \\
 HSE06\footnotemark[1]&0&0& 0.02       & -   & 0.06  &   -   \\
 HSEsol\footnotemark[1] \footnotemark[2]& 0 & 0 &0.01& - &  0.04   & - \\
\end{tabular}
\end{ruledtabular}
\footnotetext[1]{Estimated with fixed geometry obtained from SCAN\{6, 12\} relaxation}
\footnotetext[2]{No D3 correction}

    \label{table:int_int}
    \end{center}
\end{table}

We have computed classical diffusion kinetic parameters (activation energies and vibrational prefactors) for diffusion hops between hydrogen interstitial sites using a hybrid search approach. Here, candidate transition state configurations were identified using Climbing-Image Nudged
Elastic Band (CI-NEB) calculations\cite{henkelman2000climbing} with a maximum-force criterion of $0.05 \ \mathrm{eV/ \AA}$. Further refinement of our transition state search was performed using the dimer method\cite{henkelman1999dimer} and a $0.01 \ \mathrm{eV/ \AA}$ maximum-force criterion. Similar to our previous analysis, the normal mode harmonic frequencies for all transition state saddle points were obtained using dynamical matrix calculations with a displacement magnitude of $10^{-3} \ \mathrm{\AA}$. HSEsol and HSE06 transition energies were estimated from single point calculations from the SCAN\{6, 12\} result. Transitions involving the int and lin sites are likely more relevant than those with oct sites since absorption in the latter is significantly less energetically favorable. Hence, we initially focus on on int/lin diffusion hops in our work.

We have determined that transitions between neighboring int sites involves a three-step reaction, involving a rotation from an int site into the nearest lin site (int-lin), diffusion along the O--O axis to the adjacent lin site where the hydrogen ion is now bonded to the nearest-neighbor oxygen site (lin-lin), and finally a second rotation to the neighboring int site (lin-int; Fig.~\ref{fig:int_int_configs}). We find that the entire set of our DFT calculations yields reaction barriers on the order of 0.1~eV or less for all steps in this reaction pathway (Fig.~\ref{fig:int_int_rxn_coord} and Table~\ref{table:int_int}). All results for the int-lin activation energy from PBE, PBEsol, and SCAN range from values of $0.03 - 0.06$~eV, with little to no dependence on $\mathrm{U^{O_{2p}}_{eff}}$. HSE06 and HSEsol yielded estimated barriers of 0.02 and 0.01~eV, respectively. 

Separately, our computed values for the lin-lin transition exhibited a slightly broader range of results. The largest barriers resulted from setting $\mathrm{U^{O_{2p}}_{eff}} = 12$, though this effect was small. For example, PBEsol$\mathrm{\{6, 0\}}$ yielded an activation energy of 0.05~eV, whereas PBEsol$\mathrm{\{6, 12\}}$ yielded a value of 0.07~eV. Results from the PBEsol set were slightly lower, where PBEsol$\mathrm{\{6, 0\}}$ yielded a barrier of 0.02~eV and PBEsol$\mathrm{\{6, 12\}}$ a value of 0.04~eV. Results from the SCAN set were slightly higher, where SCAN$\mathrm{\{6, 0\}}$ yielded a value of 0.10~eV and SCAN$\mathrm{\{6, 12\}}$ a value of 0.14~eV. For comparison, we estimate values of 0.06 and 0.04~eV from HSE06 and HSEsol, respectively. For our calculations, all of the diffusion hop barriers discussed thus far have high likelihood for activation at ambient temperatures ($\sim$0.03~eV). 

Our computed kinetic prefactors vary strongly according to reaction type, functional, and choice of Hubbard U parameters (see Appendix \ref{sec:prefactors}), in part due to the difficulty in numerically resolving these values.
We note that hydrogen can diffuse through the CeO$_2$ lattice via int and lin sites, only, which implies that hydrogen absorbed in CeO$_2$ is potentially highly diffusive under many experimental conditions. 
Our range of results are comparable to the bulk diffusion energy barrier reported by Ref.\citenum{wu2015role} ($0.18 \ \mathrm{eV}$). The moderate differences in energetic barriers from our results can be explained by their use of a different functional (PW91), smaller basis set (400 eV), and from system size effects due to smaller lateral supercell dimensions in their slab configuration. 
Results from Ref.~\citenum{marrocchelli2012first} report a diffusion barrier of 0.52~eV, though those surface slabs calculations include adsorbed S atoms and use the PW91 functional and 400~eV basis set. In addition, the diffusion hopping path in Ref.~\citenum{marrocchelli2012first} is not clearly defined.

To estimate the significance of quantum mechanical tunneling, we now evaluate the characteristic crossover temperatures (i.e., the temperature at which the probability for thermally driven, over-barrier hopping equals that of tunneling) using the following equation:\cite{gillan1987quantum}

\begin{equation}
T_c = \frac{h |\nu^*|}{2 \pi k_B}.
\end{equation}

\noindent Here, $\nu^*$ is the imaginary frequency at the transition state, $h$ is Planck's constant, $k_b$ is Boltzmann's constant, and $T_c$ is the crossover temperature for a parabolic barrier. 
$T_c$ values for all barriers and non-HSE functionals are included in Table \ref{table:int_int}. 
The values for the int-lin transitions are $\sim 55 \ \mathrm{K}$ with little variation, suggesting that those transitions are primarily driven by over-barrier diffusion at ambient conditions.
The $T_c$ values for the lin-lin transition range from $80 - 348 \ \mathrm{K}$, where only SCAN$\mathrm{\{6, 6\}}$ and SCAN$\mathrm{\{6, 12\}}$ exhibit above ambient values (298~K), with crossover temperatures of 313 and 348~K, respectively. 
This result is expected considering that SCAN yields the largest lin-lin energy barriers as well as the stiffest $\nu \mathrm{(OH)}$ values.
Our results suggest that the SCAN calculations yield an upper-bound on the hydrogen diffusion activation energy, which is likely effectively decreased at low temperatures due to tunneling effects. 

The energetic barriers and rate prefactors for the oct-int transition are included for the sake of completeness, despite the large oct site absorption energy (Appendix \ref{sec:oct-int}). All three sets of calculations with PBE, PBEsol, and SCAN indicate an inverse correlation with $\mathrm{U^{O_{2p}}_{eff}}$, where $\mathrm{U^{O_{2p}}_{eff}}=0$ yields the highest barriers, with values of 0.03~eV, 0.04~eV, and 0.06~eV, respectively. However, the $\mathrm{U^{O_{2p}}_{eff}} = 12$ results are only marginally smaller, with values of 0.01~eV, 0.02~eV, and 0.02~eV, for PBE, PBEsol, and SCAN, respectively. The activation energy for the reverse transition (int-oct) is $\sim$1~eV higher, given that the formation energies for oct site absorption is approximately that much greater than that of the int site.  
We estimate barrierless transitions (negative energy barriers) for HSE06 and HSEsol, though this is likely due to poor accuracy of the single-point, fixed-atomic-geometry estimates used for those calculations. 

\section{Conclusions}

In this work, we have utilized an array of DFT calculations with different functionals and options to correct for self-interaction in order to place reasonable bounds on hydrogen diffusion in the bulk of stoichiometric $\mathrm{CeO_2}$. Our survey includes results from the PBE and PBEsol GGA functionals, the SCAN meta-GGA functional, as well as the HSE and HSEsol hybrid functionals. For non-hybrid functionals, we applied a Hubbard $\mathrm{U_{eff}}$ value of $6 \ \mathrm{eV}$ to the Ce 4$f$ states and tested $\mathrm{U_{eff}}$ for the O 2$p$ states over a range of values from $0-12 \ \mathrm{eV}$. We have evaluated each set of DFT parameters by computing the $\mathrm{CeO_2}$ lattice constant, band gaps, bulk modulus, hydrogen absorption energies, O--H bond vibration frequency, and kinetic parameters of several interstitial hydrogen diffusion pathways. 

For bulk, defect-free $\mathrm{CeO_2}$ properties, we find that non-zero values of $\mathrm{U^{O_{2p}}_{eff}}$ gave modest improvements to the accuracy of lattice constants and generally improved the O2$p$-Ce4$f$ and the O2$p$-Ce5$d$ band gaps, though the effect on the O2$p$-Ce5$d$ gap was somewhat smaller. We observe that the predicted bulk modulus generally increased monotonically with increasing values of $\mathrm{U^{O_{2p}}_{eff}}$. However, all values excluding those from PBE are within the experimental range of results. We find a stronger functional dependence for these properties, with the SCAN and PBEsol calculations achieving results most consistent with experimental values for both the band gaps and bulk modulus.  

We find the effect of $\mathrm{U^{O_{2p}}_{eff}}$ on hydrogen absorbed within the $\mathrm{CeO_2}$ lattice to be less significant, where the variability in O--H bond frequencies, H absorption energies, and kinetic parameters for the different tested values was generally minor. In contrast, these properties showed a stronger dependence on the choice of functional. We determined an O--H bond vibrational frequency of $\sim$3490 cm$^{-1}$ from PBE at the low end of our survey, compared to a value of $\sim$3630 cm$^{-1}$ from SCAN at the high end and the experimentally observed frequency at 3510 $\mathrm{cm}^{-1}$. \cite{badri1996ftir} Across the DFT approaches used here, the ``linear'' (lin) interstitial site most often yielded the lowest formation energy, with the results ranging from values of $\sim$0.9~eV from PBE to $\sim$1.4~eV from the HSE functionals. Generally, the ``interstitial'' (int) site was only a few hundredths of an electron volt greater. Finally, the octahedral (oct) site was consistently less energetically favorable for all functionals, with absorption energies ranging from $\sim$2.5~eV from PBE to $\sim$3.3~eV from HSE06.  

Overall, we find that H diffusion in the bulk of stoichiometric $\mathrm{CeO_2}$ is likely activated at ambient temperatures, with an upper bound on the activation energy barrier of 0.14~eV from SCAN. This diffusion rate from SCAN could be enhanced at ambient temperature by hydrogen quantum tunneling, though in general we find that are computed diffusion hop barriers are low enough that quantum vibrational effects are likely minimal. Our comprehensive study could help future DFT Ce-O-H studies by narrowing down the DFT parameter space to investigate. Furthermore, the kinetic results of this work can act as a reasonable basis for parameterizing coarse-grained models that could help elucidate future experiments.  

\section{Acknowledgements}

We would like to acknowledge informative discussions with Mark Burton, Matthew Kroonblawd, Nathan Keilbart, and Sebastien Hamel. This work was performed under the auspices of the U.S. Department of Energy by Lawrence Livermore National Laboratory under Contract DE-AC52-07NA27344.

\section{Author declarations}
The authors have no conflicts to disclose. 

\section{Author contributions}
Jared Stimac: Formal analysis (lead); Writing/Review \& Editing (equal).
Nir Goldman: Conceptualization (lead); Writing/Review \& Editing (equal); Formal analysis (supporting).

\section{Data Availability}
The data that supports findings of this work can be made available upon reasonable request from the corresponding author. 

\bibliography{references}

\clearpage

\appendix

\section{Rate prefactors for transitions between neighboring int sites}
\label{sec:prefactors}

The classical prefactor rates were calculated using the harmonic frequencies: 

\begin{equation}
A = \frac{\prod_{i}^{N} \nu_i^{min} }{\prod_j^{N-1} \nu_j^{\ddag}}
\label{eqn:prefactor}
\end{equation}

where $\nu_i^{min}$ and $\nu_j^{\ddag}$ are the frequencies at a barrier's energy minima and transition state, respectively.

\begin{table}[hbt!]
    \caption{Transition rate prefactors $A$. Values not included reflect  calculations that did not converge with increasing degrees of freedom.}
    \begin{center}
    \include{prefactors_int_int_table}
    \end{center}
    \label{table:prefactors}
\end{table}

\clearpage

\section{Oct-int barrier}
\label{sec:oct-int}
\begin{table}[hbt!]
    \caption{Oct to int transition energy barrier $\Delta E$, rate prefactor $A$, and crossover temperature $T_{\mathrm{c}}$}
    \begin{center}
    \begin{ruledtabular}
\begin{tabular}{cccccc}

      & \multicolumn{2}{c}{$\mathrm{U_{eff}}$ [eV]}         &              &             &           \\[-.1cm]
$\mathrm{E_{XC}}$& Ce 4\textit{f}    &  O 2\textit{p}       &  $\Delta E$ [eV]  & $A$ [THz] &    $T_\mathrm{c}$ [K] \\
\hline
PBE     & 6 &        0 &    0.03  &  1.8     & 100  \\
        & 6 &        6 &    0.02   &   2.9    &   80  \\
        & 6 &       12 &    0.01   &    1.6    &   66  \\
PBEsol  & 6 &        0 &    0.04   &   2.3    &  104  \\
        & 6 &        6 &    0.03   &    1.6    &   96  \\
        & 6 &       12 &    0.02   &   1.5    &   78  \\
SCAN    & 6 &        0 &    0.06   &   2.5    &  114  \\
        & 6 &        6 &    0.04   &    2.0    &  115  \\
        & 6 &       12 &    0.02   &    3.2    &   90  \\
HSE06\footnotemark[1]& 0 &        0 &  -0.01      &      -     &    - \\
HSEsol\footnotemark[1] \footnotemark[2] & 0 &        0 &    -0.01     &    -    &  -   \\

\end{tabular}
\end{ruledtabular}
\footnotetext[1]{Estimated with fixed geometry obtained from SCAN\{6, 12\} relaxation}
\footnotetext[2]{No D3 correction}
    \end{center}
    \label{table:oct_int}
\end{table}

\clearpage

\section{Convergence}
\label{sec:Convergence}
To further quantify the convergence the DFT calculations, this section includes additional testing regarding the energy activation energy barrier and the classical prefactors. 

First, the convergence of the oct-int SCAN$\{6,\ 6\}$ activation energy barrier was tested with single-point calculations (atomic nuclei fixed) from the CI-NEB result, prior to implementing the dimer method. The density of the k-point mesh and the plane-wave cutoff were varied independently. Increasing the mesh density from $2\times2\times2$ to $4\times4\times4$ resulted in a deviation of magnitude  $3\times 10^{-3} \ \mathrm{eV}$. Increasing the plane-wave cutoff from $500 \ \mathrm{eV}$ to $600 \ \mathrm{eV}$ resulted in a deviation of magnitude  $1\times 10^{-2} \ \mathrm{eV}$.

Now considering the convergence of the classical prefactors, $A$, that were estimated using the harmonic frequencies evaluated with dynamical matrix calculations. 
The most substantial and illusive source of perturbation effecting estimations of $A$ likely come from metastability brought on by the Hubbard U correction.
This explains several instances in which the calculated $A$ values diverged to nonphysical magnitudes with increasing degrees of freedom ($N_{\mathrm{dof}}$). 
Consider Table \ref{table:prefactor_dof}, which presents $A$ at different $N_{\mathrm{dof}}$ for SCAN$\{6,\ 6\}$(converged) and SCAN$\{6,\ 12\}$(diverged).

\begin{table}[hbt!]
    \caption{Convergence of the SCAN$\{6,\ 6\}$ and SCAN$\{6,\ 12\}$ int-lin transition rate prefactors $A$ with degrees of freedom $N_{\mathrm{dof}}$ included in dynamical matrix calculation. Increasing $N_{\mathrm{dof}}$ results from inclusion of atomic positions within further radii of the H atom.}
    \begin{center}
    \begin{ruledtabular}
\begin{tabular}{ccc}
 
$\mathrm{U_{eff}}$(O2$p$) & $N_{\mathrm{dof}}$ & $ A \ \mathrm{[THz]}$  \\
\hline
6    &   24   & 10.0 \\
6    &   165  & 8.2  \\
6    &   291  & 7.8 \\
\hline
12   &   24   & 5.6 \\
12   &   165  & 70.9  \\
12   &   291  & 2201.8 \\
\end{tabular}
\end{ruledtabular}

    \end{center}
    \label{table:prefactor_dof}
\end{table}

\begin{figure}
    \includegraphics[width=1.0\textwidth]{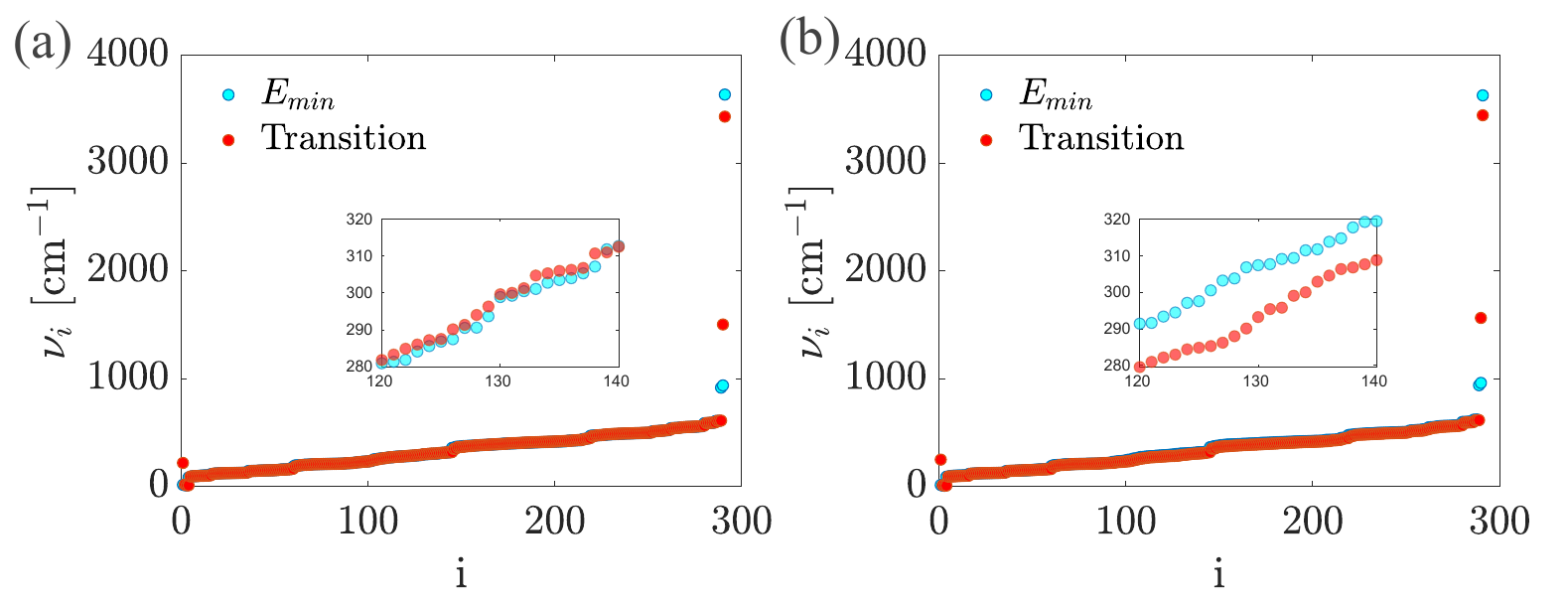}
    \caption{Frequency distributions $\nu_i$ for (a) SCAN$\{6,\ 6\}$, and (b) SCAN$\{6,\ 12\}$. Both figures depict frequencies at the energy minimum $E_{min}$ and the transition. The frequencies are listed in ascending order, but starting with the few that are imaginary (one expected at the transition and three smaller noise-associated transnational modes.}
    \label{fig:Freq_dist}
\end{figure}

This issue seems to perturb any individual frequency only by a small degree. 
This is made evident when comparing the distribution of harmonic frequencies for the SCAN$\{6,\ 6\}$ and SCAN$\{6,\ 12\}$ for the int-lin transition, as shown in Fig. \ref{fig:Freq_dist}. 
That is, the overall distributions are quite similar, basically indistinguishable viewing the full distribution; its only when we examine some of the frequencies more closely, as done in the figure inset, that difference is noticeable. 
These two subfigures would not be expected to have the same exact frequency distribution, as they result from different Hubbard corrections, but they should be similar.
Notice in the inset of SCAN$\{6,\ 12\}$   that energy minimum frequencies are $\sim 5\%$ above those at the transition and energy minimum configuration frequencies, which is not found in the SCAN$\{6,\ 6\}$.
The uncertainty believed to be caused metastabilities has a dramatic effect on a subset of the calculations, but its unclear if those calculations that converged to physically plausible prefactor magnitudes were still affected to some degree.

Although the metastability issue limits our confidence in the estimated prefactor values, we also looked at their convergence with other computational parameters for a case in which metastabilities did not lead to nonphysical divergence.   
In Table \ref{table:A_convergence_1}, the prefactor for the PBE$\{6,\ 6\}$ int-lin barrier for different displacement magnitudes used in the dynamic matrix calculation as well as a different convergence criterion of the maximum force ($f_{\mathrm{max}}$) used to obtain the energy minimum and the transition configurations.

\begin{table}[hbt!]
    \caption{ Rate prefactor $A$ for various displacement magnitudes used for the dynamical matrix calculation and maximum force ($f_{\mathrm{max}}$) used to obtain transition and energy minimum configurations. }
    \begin{center}
    \begin{ruledtabular}
\begin{tabular}{ccc}
 
Displacement $\mathrm{[\AA]}$ & $f_{\mathrm{max}} \ \mathrm{[eV/\AA]}$ & $ A \ \mathrm{[THz]}$  \\
\hline
$5 \times 10^{-3}$     & $1 \times 10^{-1}$ &   24.0 \\
$2 \times 10^{-3}$     & $1 \times 10^{-1}$ &   4.1  \\
$1 \times 10^{-3}$     & $1 \times 10^{-1}$ &   7.2  \\
$5 \times 10^{-4}$     & $1 \times 10^{-1}$ &   1.6\footnotemark[1]  \\
\hline
$1 \times 10^{-3}$     & $1 \times 10^{-4}$ &   10.1 \\

\end{tabular}
\end{ruledtabular}
\footnotetext[1]{Notable increase in the magnitude in the three imaginary transitional modes that are filtered out to evaluate $A$, including one imaginary frequency at the energy minimum configuration that exceeded $100 \ \mathrm{cm^{-1}}$.}
    \end{center}
    \label{table:A_convergence_1}
\end{table}

\end{document}